# Using Categorization of Problems as an Instructional Tool to Help Introductory Students Learn Physics

Andrew Mason[1] and Chandralekha Singh[2]

1. Department of Physics and Astronomy, University of Central Arkansas, Conway, AR 72035
2. Department of Physics and Astronomy, University of Pittsburgh, Pittsburgh, PA 15260

## ABSTRACT

The ability to categorize problems based upon underlying principles, rather than contexts, is considered a hallmark of expertise in physics problem solving. With inspiration from a classic study by Chi, Feltovich, and Glaser, we compared the categorization of 25 introductory mechanics problems based upon similarity of solution by students in large calculus-based introductory courses with physics faculty and Ph.D. students. Here, we summarize the study and suggest that a categorization task, especially when conducted with students working with peers in small groups, can be an effective pedagogical tool to help students in introductory physics courses learn to discern the underlying similarity between problems with diverse contexts but the same underlying physics principles.

## INTRODUCTION

Physics is a knowledge-rich subject, and the laws of physics are encapsulated in precise mathematical forms. Students in many introductory physics courses must learn to unpack the compact mathematical laws of physics and apply them in diverse situations to explain and predict physical phenomena.[1-4] In other words, in order to become an expert, students must learn to interpret and make sense of abstract physical principles and make a conscious effort to build a coherent knowledge structure.[5-9] Categorizing or grouping together physics problems based upon similarity of solution, instead of contexts or "surface features" of the problems, is considered a hallmark of expertise.[10-14] An expert in physics may categorize many problems involving conservation of energy in one category and those involving conservation of momentum in another category, even if some of the problems involving the different conservation laws may have similar contexts and other problems involving conservation of energy alone may have very different contexts.

However, learning is context dependent, and many students in introductory physics courses struggle to discern the underlying similarity of physics problems with different surface features. For example, a physics professor will group together a problem involving finding the speed of a spinning ballerina who puts her arms close to her body and a problem involving spinning neutron star that is collapsing under its own gravitational force as similar, because in both problems, there is no external torque on either system, and the conservation of angular momentum of each system implies that the angular speed increases. However, students may focus on the surface features of the problem, and may view the ballerina and neutron star problems as very different. In the classic study conducted by Chi, Feltovich and Glaser[10] (here called the Chi study), eight introductory physics students in calculus-based courses were asked to categorize introductory mechanics problems based upon similarity of solution. Unlike experts, who categorize problems based on the physical principles involved in solving them, introductory students were sensitive to the surface features of the problems – for example, placing problems involving inclined planes in one category and problems involving pulleys in a separate category.[10]

With inspiration from the Chi study[10], we compared the categorization of introductory mechanics problems by students in large calculus-based courses with physics faculty and Ph.D. students. We find a significantly wider spectrum in students' expertise in large introductory classes than was captured by analyzing data from only 8 introductory student volunteers in the Chi[10] study. In the following sections, we describe the study and suggest that a categorization task, especially when conducted with students working with peers in small groups, can be an effective pedagogical tool to help students in introductory physics courses learn to discern the underlying similarity between problems with diverse contexts but the same underlying physics principles.

## METHODOLOGY

We were unable to obtain the questions in Chi et al.'s study.[10] We therefore chose our own mechanics questions and many questions were adapted from previous studies.[11,13] The context of the 25 mechanics problems varied and the topics included one- and two-dimensional kinematics, dynamics, work-energy, and impulse-momentum.[15] All participants who performed the categorization task were provided the following instruction at the beginning of the problem set:

*Your task is to group the 25 problems below based upon similarity of solution into various groups on the sheet of paper provided. Problems that you consider to be similar should be placed in the same group. You can create as many groups as you wish. The grouping of problems should NOT be in terms of ``easy problems'', ``medium difficulty problems'' and ``difficult problems'' but rather it should be based upon the features and characteristics of the problems that make them similar. A problem can be placed in more than one group created by you. Please provide a brief explanation for why you placed a set of questions in a particular group. You need NOT solve any problems.*

The sheet on which participants were asked to perform the categorization of problems had three columns. The first column asked them to use their own category name for each of their categories, the second column asked them for a description of each category that explains why problems within that category may be grouped together, and the third column asked them to list the problem numbers for the questions that should be placed in a category. Apart from these directions, neither students nor faculty were given any other hints about which category names they should choose.

Although we had an idea about which categories created by individuals should be considered good or poor, we validated our assumptions with professors. We randomly selected the categorizations performed by twenty introductory physics students, gave them to three professors who had taught introductory physics recently, and asked them to decide whether each of the categories created by individual students should be considered good, moderate, or poor. We asked them to mark each row which had a category name created by a student and a description of why it was the appropriate category for the questions that were placed in that category. If a faculty member rated a category created by an introductory student as good, we asked that he/she cross out the questions that did not belong to that category. The agreement between the ratings of different faculty members was better than 95%.

We used their ratings as a guide to rate the categories created by everybody as good, moderate, or poor. A category was considered "good" only if it was based on the underlying physics principles. We typically rated both conservation of energy and conservation of mechanical energy as good categories. Kinetic energy as a category name was considered a moderate category if students did not explain that the questions placed in that category could be solved using mechanical energy conservation or the work energy theorem. We rated a category such as energy as good if students explained the rationale for placing a problem in that category. If a secondary category such as friction or tension was the only category in which a problem was placed and the description of the category did not explain the primary physics principles involved, it was considered a moderate category. Categories such as ramps and pulleys were rated as poor.

More than one principle or concept may be useful for solving a problem. The instruction for the categorizations told students that they could place a problem in more than one category. Because a given problem can be solved using more than one approach, categorizations based on different methods of solution that are appropriate were considered good. For questions that required the use of two major principles, those who categorized them in good categories either made a category which included both principles, such as the conservation of mechanical energy and the conservation of momentum, or placed such questions in two categories created by them: one corresponding to the conservation of mechanical energy and the other corresponding to the conservation of momentum. If such questions were placed only in one of the two categories, the categorization was considered not good. For some questions, conservation of mechanical energy may be more efficient, but the questions can also be solved using one- or two-dimensional kinematics for constant acceleration. Here, we will only discuss categories that were rated good by faculty members. In the results section below, if a particular group (introductory students, Ph.D. students, or professors) placed 60% of the questions in a good category, it is implied that the other 40% of the questions were placed in moderate or poor categories.

## RESULTS

A calculus-based class with 180 students performed the categorization in their recitation classes after relevant instruction. Figure 1 shows a histogram of the percentage of questions placed in good categories (not moderate or poor) and compares the average performance on the categorization task of 180 introductory students with 21 Ph.D. students and 7 physics faculty. Although the categorization by the calculus-based introductory students is not on par with the categorization by physics Ph.D. students, there is a large overlap between the two groups, which suggests that many calculus-based introductory students are far from being "novices" when categorizing the problems.[10]

Figure 1 also shows that the difference in good categorization performed by the physics professors and physics Ph.D. students is much larger than the difference between Ph.D. students and the calculus-based introductory physics students. Physics professors often pointed out multiple methods for solving a problem and specified multiple categories for a particular problem more often than the Ph.D. students and introductory students. Introductory physics students and even some Ph.D. students were much more likely to place questions in inappropriate categories than the professors, e.g., placing a problem that is based on the impulse-momentum theorem or conservation of momentum in the conservation of energy category.

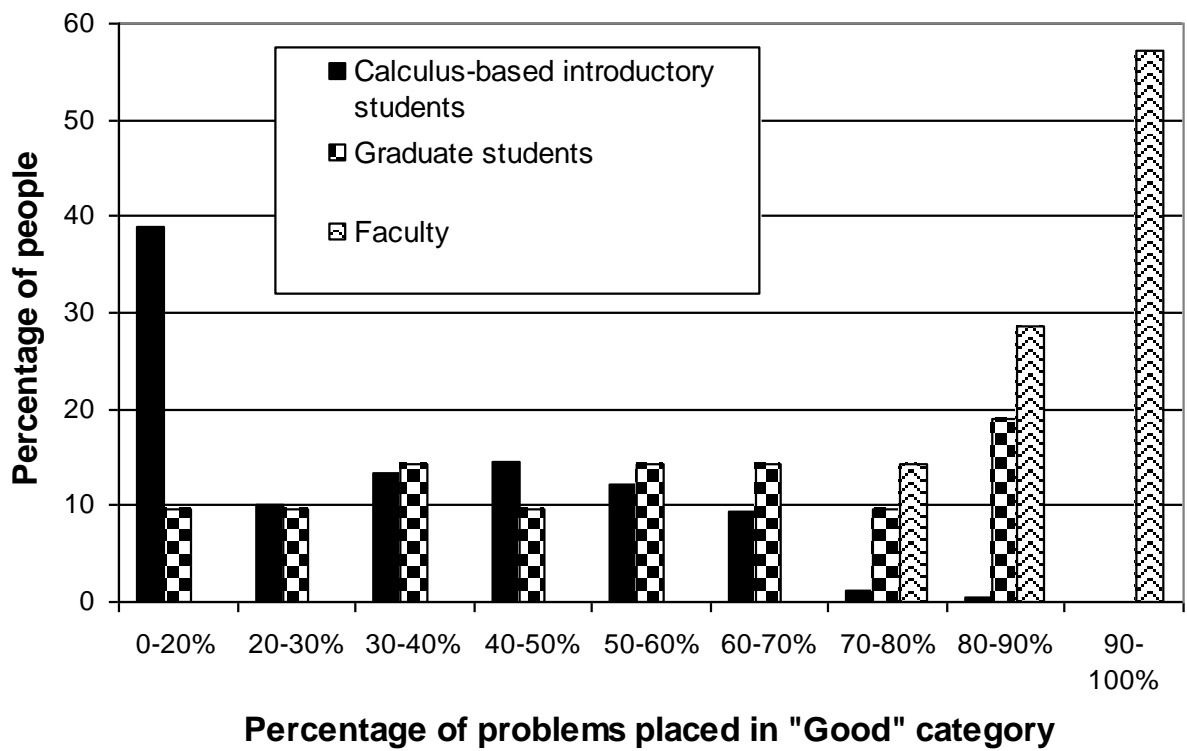

**Figure 1. Histogram of calculus-based introductory physics students, Ph.D. students, and physics faculty who categorized various percentages of the 25 problems in version I in "good" categories when asked to categorize them based on similarity of solution. Physics faculty members performed best in the categorization task. There is a large overlap between the Ph.D. students and the introductory physics students.**

Many of the categories generated by the faculty, Ph.D. students and introductory students were the same, but there was a difference in the fraction of questions that were placed in good categories by each group. There were some poor categories such as "ramps" and "pulleys" that were made by introductory physics students but not by physics faculty or Ph.D. students. Moreover, some introductory physics students created the categories such as speed and kinetic energy if the question explicitly asked them to calculate those physical quantities. The explanations provided by the students as to why a particular category name, for example, speed, is most suitable for a particular problem were not adequate; they wrote that they created this category because the question asked for the speed. As expected, Ph.D. students were less likely than introductory students to create such categories and were more likely to classify questions based on physical principles and concepts, for example, conservation of mechanical energy or kinematics in one dimension.

In questions involving application of two major physics principles, for example, in the ballistic pendulum problem, most faculty categorized the problem in both "conservation of mechanical energy" and "conservation of momentum" categories. In contrast; most introductory students and even some Ph.D. students either categorized it as an energy problem or as a momentum problem but not both.

## IMPLICATION FOR INSTRUCTION

Teaching introductory physics using diverse contexts is a useful tool for helping students learn physics. Students must learn to apply physics principles in diverse situations to explain and predict physical phenomena. The concrete contexts can help students learn abstract physics principles better. The contexts can also help keep students actively engaged in the learning process and help them connect what they are learning with their prior knowledge and experiences. However, students also need to learn to see beyond a particular context to the underlying physics principles, and understand why those principles are applicable in those contexts and how they would know if a particular principle would be applicable in a new situation they encounter.

Categorizing various problems based on similarity of their solutions can be a useful tool to help students learn physics, because such tasks can guide students to focus on the similarity of problems based on the underlying principles rather than on the specific contexts. For example, introductory physics students with different levels of expertise can be placed in small groups, in which they are asked to categorize problems and discuss why different problems should be placed in the same group, without asking them to solve the problems explicitly. Since there is diversity in the individual performance on categorization task as shown in Figure 1, introductory students can have meaningful discussions while categorizing problems in small groups. Then, there can be a class discussion about why some categorizations are better than others, and students can be given a follow-up categorization task to ensure individual accountability in learning from this activity. Categorization tasks help students practice underdeveloped skills; they only perform the conceptual analysis and planning of the problem solution in the problem solving process as opposed to implementing the plan, which discourages algorithmic approach to problem solving. Also, instructors can develop their own pedagogical content knowledge in terms of common student difficulties in categorization of problems, by categorizing questions from the perspective of their students, and then comparing their responses to published data about how introductory students actually categorize these problems when working alone.[11]

## ACKNOWLEDGEMENTS

We thank the US National Science Foundation for award PHY-1202909.